\documentstyle[12pt]{article} \textwidth=18cm \oddsidemargin=-1cm
\begin{document} \centerline{\bf
VIOLATION OF THE GAUGE EQUIVALENCE}

R. I. Khrapko\footnote{Moscow Aviation Institute, 125871, Moscow,
Russia, khrapko\_ri@hotmail.com} \medskip

\centerline{\bf Abstract } \medskip

F. V. Gubarev et al. [4] have argued that the vector potential itself
may have physical meaning, in defiance of the gauge equivalence
principle.  Earlier, R. I.  Khrapko proposed a gauge noninvariant
electrodynamics spin tensor [1]. The standard electrodynamics spin
tensor is zero.

Here we point out that the Biot-Savarat formula uniquely results in a
preferred, ``true" vector potential field which is generated from a
given magnetic field. A similar integral formula uniquely permits to
find a ``true" scalar potential field generated from a given electric
field even in the case of a nonpotential electric field.

A conception of differential forms is used. We say that an exterior
derivative of a form is the boundary of this form and the integration
of a form by the Biot-Savarat-type formula results in a new form named
the {\it generation}. Generating from a generation yields zero. The
boundary of a boundary is zero. A boundary is closed. A generation is
{\it sterile}. A {\it conjugation} is considered. The conjugation
converts closed forms to sterile forms and back. It permits to construct
chains of forms. The conjunction differs from the Hodge star operation:
the conjugation does not imply the dualization. A circularly polarized
wave is considered in view of the electrodynamics spin tensor problem.

A new anthropic principle is presented.
\medskip

{\bf 1. The gauge equivalence of differential forms } \medskip

It is obvious that in a static case we can add a constant $\phi_0$ to
an electric scalar potential $\phi$ and we can add a gradient
$\partial_if$ to a magnetic vector potential $A_i$ without changing the
corresponding electric $E_i$ and magnetic $B_{ij}$ fields. Indeed,
$$E_i=\partial_i\phi=\partial_i (\phi+\phi_0),\quad
B_{ij}=2\partial_{[i}A_{j]}=2\partial_{[i}(A_{j]}+\partial_{j]}f).$$

The change
$$\phi\to\phi+\phi_0,\qquad A_i\to A_i+\partial_if $$
is referred to as the {\it gauge transformations} of the potentials.
Thus, pairs of potentials $\phi,\ A_i$ which are connected by the gauge
transformations give the same electromagnetic field $E_i,\ B_{ij}$.

All these quantities are differential forms. We name them
simply {\it forms}. All derivatives considered here are external
derivatives. We say that an external derivation of a form results in
the {\it boundary} of this form, and we name the form under derivation
a {\it filling} of the boundary. So $E_i$ is the boundary of the form
$\phi$ and $B_{ij}$ is the boundary of the form $A_j$. $\phi$ is the
filling of the form $E_i$ and $A_j$ is the filling of the form
$B_{ij}$.

The forms $\phi_0$ and $\partial_if$ are referred to as {\it closed}
forms because their external derivatives are equal to zero:
$\partial_i\phi_0=0$, $\partial_{[i}\partial_{j]}f=0$. When necessary,
we mark closed forms by the symbol bullet $\bullet$:
${\mathop\phi\limits_\bullet}{}_0,\
{\mathop\partial\limits_\bullet}{}_jf$.

It is obvious that any boundary is closed. That is, the boundary of a
boundary is equal to zero. For example,
$$\partial_{[k}{\mathop E\limits_\bullet}{}_{i]}=
\partial_{[k}\partial_{i]}\phi=0,\quad
\partial_{[k}{\mathop B\limits_\bullet}{}_{ij]}=
2\partial_{[k}\partial_iA_{j]}=0.$$

The boundary of a form is determined uniquely, a filling of a form
admits an addition of a closed form. Thus, an electric field strength
$E_i$ and a magnetic induction $B_{ij}$ do not change when closed forms
are added to the potentials:
$$\phi\to\phi+{\mathop\phi\limits_\bullet},\qquad A_i\to A_i+{\mathop
A\limits_\bullet}{}_i. $$\medskip

{\bf 2. True potentials } \medskip

But, from our point of view, it does not mean that different potentials
which are connected by the gauge transformations are completely
equivalent to one another. There are preferred, "true" potentials, which
correspond to a given electromagnetic field. Early we have stated an
assumption that a spin tensor of electromagnetic waves is expressed in
terms of such a vector potential [1, 2]. The standard electrodynamics
spin tensor is zero.

A true vector potential can be get by the Biot-Savarat formula:
$$A_j(x)= \int\frac{B_{i'j}(x')r^{i'}(x,x')dV'}
{4\pi r^3(x,x')}.\eqno(1)$$
A true scalar potential can be get by a similar formula:
$$\phi(x)=\int\frac{E_{i'}(x')r^{i'}(x,x')dV'}
{4\pi r^3(x,x')}.\eqno(2)$$
We use marked indexes. Primes belong to indexes,
but not to kernel letters. In the integrals (1), (2) the primes mark a
varying point $x',$ but not another coordinate system.

The formula (2) does not meet in literature.  It determines the
potential $\phi$ uniquely, in particular, in the case of a nonpotential
electric field.

We say that $E_i$ {\it generates} $\phi$ by the formula (2) and
$B_{ij}$ generates $A_i$ by the formula (1), i.e. we say that $\phi$ is
the {\it generation from} $E_i$ and $A_i$ is the generation from
$B_{ij}$.  Otherwise, we say that $E_i$ is a {\it source} of $\phi$ and
$B_{ij}$ is a source of $A_i$. The symbol {\it dagger} $\dag$ is used
for a brief record of generating, and the symbol {\it times} $\times$
marks a generation.  So, the true potentials are given by the formulas:
$$\mathop\phi\limits_\times=\dag^iE_i,\quad
{\mathop A\limits_\times}{}_j=\dag^iB_{ij}.$$

Thus, $\dag$ is an operation which is inverse to an external
derivation. Without indexes this looks as follows:
$\dag\partial=\partial\dag=1,$ or rather $\dag\partial\dag=\dag,\
\partial\dag\partial=\partial$ (see Sec. 3)\medskip

{\bf 3. Generations are sterile } \medskip

At this point, a problem arises. What shall we get if a generation will
be used as a source of a generation? What shall we get if a generation
will be substituted in the integral formula? For example, what is the
value of the integral
$$\int\frac{{\mathop A\limits_\times}{}_{j'}(x')r^{i'}(x,x')dV'}
{4\pi r^3(x,x')}\ \mbox{?}$$

The question is a simple one: generating from a generation yields zero
[3]. We say that generations are {\it sterile}. For example,
$$\int\frac{{\mathop A\limits_\times}{}_{j'}(x')r^{i'}(x,x')dV'}
{4\pi r^3(x,x')}=0,\qquad\mbox{briefly: }\dag^i\dag^jB_{ij}=0.$$
It implies that a sterile addition to a source does not change the
generation.

Thus, \begin{center}
$\partial\ $(filling) = $\partial\ $(filling + closed
form) = boundary (which is closed),\\ $\dag\ $(source) =
$\dag\ $(source + sterile form) = generation (which is sterile)
\end{center}

Now we can decompose any form into closed and sterile parts. For
example:
$$A_k={\mathop A\limits_\bullet}{}_{k}+{\mathop A\limits_\times}{}_{k}=
\partial_k\dag^jA_j+\dag^i2\partial_{[i}A_{k]},$$
because one can name the expressions
$${\mathop A\limits_\bullet}{}_{k}=
\partial_k\dag^jA_j\quad\mbox{and }\quad
{\mathop A\limits_\times}{}_{k}= \dag^i2\partial_{[i}A_{k]}$$
closed and sterile components of the form $A_k$, respectively.
\pagebreak

{\bf 4. The conjugation. Chains of fields } \medskip

In a metric space there is a relations between contra- and covariant
tensors of the same valence (with the same number of indices).  For
example, the metric tensor $g_{ik}$ associates a tensor $X^{ij}$ with
the tensor $X_{mn}=X^{ij}g_{im}g_{jn}.$ This process is called the
lowering of indices. In this case the same kernel letter for the
quantity is used.

In the electromagnetism a slightly different process is used. We call
this process the {\it conjugation}. The conjugation establishes a
one-to-one correspondence between forms and contravariant tensor
densities.
This process uses the metric tensor densities
$g^\wedge_{ij}=g_{ij}/\sqrt{g}_\wedge$ or
$g^{ij}_\wedge=g^{ij}\sqrt{g}_\wedge$. It appears that the
electromagnetic fields are conjugated in pairs:
$$E_i=D^j_\wedge g^\wedge_{ij},\quad
D^j_\wedge=E_ig^{ij}_\wedge,\quad
B_{ik}=H^{jl}_\wedge g^\wedge_{ij}g_{kl},\quad
H^{jl}_\wedge=B_{ik}g_\wedge^{ij}g^{kl}.$$

As is known, electric induction $D^j_\wedge$, the same as electric
charge density $\rho_\wedge$, electric current density $j^i_\wedge$,
magnetic strength $H^{ij}_\wedge$ are tensor densities of weight $+1$.
To emphasize this circumstance, in serious literature Gothic characters
are used for $D^j_\wedge$, $j^i_\wedge$, $H^{ij}_\wedge$. But in the
present paper we use the symbol {\it wedge} $\wedge$ as a
sub(super)script to mark a tensor density of weight $+ 1$ or $- 1$ (see
also [3]).

Tensor densities differ from tensors: the law of transformation of
density components involves the modulus of Jacobian. For example, an
electric induction is transformed according to the formula
$$D^i_\wedge=D^{i'}_{\wedge'}\partial^i_{i'}\mid\Delta'\mid.$$
Here $\partial^i_{i'}$ is the matrix of coordinates transformation:
$\partial^i_{i'}=\partial x^i/\partial x^{i'}.$
$\Delta'=\mbox{Det}(\partial^{i'}_i)$ designates the determinant of the
inverse matrix.

The kernel letters are usually changed by conjugating of
electromagnetic fields. For brevity we designate conjugating by the {\it
star } $\star$:
$$\star E_i=D^i_\wedge,\quad\star D^i_\wedge=E_i,\quad \star
B_{ij}=H^{ij}_\wedge,\quad\star H^{ij}_\wedge=B_{ij}.$$

It is remarkable that conjugating transforms sterile fields to closed
fields and back [3]. For example,
$$ \star{\mathop E\limits_\bullet}{}_i= {\mathop
D\limits_\times}{}^i_\wedge,\quad \star{\mathop
D\limits_\bullet}{}^i_\wedge= {\mathop E\limits_\times}{}_i,\quad
\star{\mathop B\limits_\bullet}{}_{ij}= {\mathop
H\limits_\times}{}^{ij}_\wedge,\quad \star{\mathop
A\limits_\times}{}_j= {\mathop A\limits_\bullet}{}^j_\wedge. $$
By tradition, we have not changed the kernel letter $A$ in the last
equality.

So, the true vector potential ${\mathop A\limits_\times}{}_j$ becomes a
closed one, ${\mathop A\limits_\bullet}{}^j_\wedge,$ by conjugating. It
implies that ${\mathop A\limits_\bullet}{}^j_\wedge$ satisfies the
Lorentz condition
$$\partial_j{\mathop A\limits_\bullet}{}^j_\wedge=0.$$

Conjugating transforms a sterile generation to a closed field, and the
new field appears to be ready for new generating. So chains of forms,
finite or infinite, arise. We present an example of an infinite chain.
$$ \dag^i{\mathop j\limits_\bullet}{}^i_\wedge= {\mathop
H\limits_\times}{}^{ij}_\wedge,\, \star{\mathop
H\limits_\times}{}^{ij}_\wedge= {\mathop B\limits_\bullet}{}_{ij},\,
\dag^i{\mathop B\limits_\bullet}{}_{ij}= {\mathop
A\limits_\times}{}_j,\, \star{\mathop A\limits_\times}{}_j= {\mathop
A\limits_\bullet}{}^j_\wedge,\, \dag^k{\mathop
A\limits_\bullet}{}^j_\wedge= {\mathop {\cal
H}\limits_\times}{}^{jk}_\wedge,\, \star{\mathop {\cal
H}\limits_\times}{}^{jk}_\wedge= {\mathop {\cal
B}\limits_\bullet}{}_{jk},\dots $$
The script characters ${\cal H}$ and ${\cal B}$ designate hypothetical
fields. These fields arise when the chain is constructed. It is another
generation.

Conjugating permits recurring derivations. So, a chain can be
constructed
in the reverse direction by external differentiation. For example:
$$ 2\partial_{[i} {\mathop A\limits_\times}{}_{j]}=
{\mathop B \limits_\bullet}{}_{ij}, \,
\star {\mathop B \limits_\bullet}{}_{ij} =
{\mathop H \limits_\times}{}_\wedge^{ij} ,\,
\partial_j{\mathop H \limits_\times}{}_\wedge^{ij} =
{\mathop j \limits_\bullet}{}_\wedge^{i},\,
\star{\mathop j \limits_\bullet}{}_\wedge^{i} =
{\mathop j \limits_\times}{}_{i} ,\,
2\partial_{[k}{\mathop j \limits_\times}{}_{i]} =
{\mathop {\mbox{\large B}} \limits_\bullet}{}_{ki} ,\,
\star\!\mathop {\mbox{\large B}} \limits_\bullet{}_{ki} =
{\mathop {\mbox{\large H}} \limits_\times}{}_\wedge^{ki} ,\dots $$
The large charactes {\large H} and {\large B} designate hypothetical
fields. These fields arise when the chain is constructed.  It is another
generation.

Conjugating makes it possible to express the operator $\nabla^2$ in
terms of the external derivatives. It appears that
$$\nabla^2\mathop\omega\limits^p=
(-1)^p(\star\partial\star\partial-\partial\star\partial\star)
\mathop\omega\limits^p,\qquad
\nabla^2{\mathop\alpha\limits^p}{}_\wedge=
(-1)^{p+1}(\star\partial\star\partial-\partial\star\partial\star)
{\mathop\alpha\limits^p}{}_\wedge.$$
Here $\mathop\omega\limits^p$ and ${\mathop\alpha\limits^p}{}_\wedge$
designate a form of the degree $p$ and a contravariant density of
valence $p$, respectively. For example,
$$\nabla^2{\mathop A\limits_\bullet}{}^i_\wedge=
-{\mathop j\limits_\bullet}{}^i_\wedge.$$

We denote the integral operator which is inverse to $\nabla^2$ by
{\it double dagger} $\ddag.$ As is known,
$$\ddag=-\int\frac{dV'}{4\pi r(x,x')}.$$

The requirement $\nabla^2\ddag=1$ yields
$$\ddag \mathop\omega\limits^p =
(-1)^{p+1}(\star\dag\star\dag-\dag\star\dag\star)
\mathop\omega\limits^p,\qquad
\ddag {\mathop\alpha\limits^p}{}_\wedge=
(-1)^{p}(\star\dag\star\dag-\dag\star\dag\star)
{\mathop\alpha\limits^p}{}_\wedge.$$
For example,
$$\ddag{\mathop j\limits_\bullet}{}^i_\wedge=
-{\mathop A\limits_\bullet}{}^i_\wedge, \qquad
\ddag{\mathop B\limits_\bullet}{}_{ij}=
-{\mathop {\cal B}\limits_\bullet}{}_{ij}.$$
\pagebreak

{\bf 5. Vector potential squared } \medskip

The article [4] is an occasion for this paper writing. The authors of
the article [4] ``argue that the minimum value of the volume integral
of $A^2$ may have physical meaning''. In other words, the potential
which minimizes the volume integral is a preferred potential. The
authors have designated such a potential $A_{\rm min}.$

We have named such a potential ``true'' potential: ${\mathop
A\limits_\times}{}_j=\dag^i2\partial_{[i}A_{k]}.$ Therefore, the
mentioned volume integral should be evaluated by the formula
$$ <{\mathop A\limits_\times}{}_j\cdot
\star{\mathop A\limits_\times}{}_j>=
<{\mathop A\limits_\times}{}_j\cdot
{\mathop A\limits_\bullet}{}^j_\wedge>=
\int{\mathop A\limits_\times}{}_j
{\mathop A\limits_\bullet}{}^j_\wedge dV^\wedge$$
($dV^\wedge$ is a density of weight $- 1$).

However, the authors use another formula:
$$<A_{\rm min}^2>=
\int\int\frac{\hbox{\bf B}(x')\cdot\hbox{\bf B}(x)dVdV'}
{4\pi r(x,x')}.$$

This formula can be obtained by transforming the expression
$ <{\mathop A\limits_\times}{}_j\cdot{\mathop
A\limits_\bullet}{}^j_\wedge> :$
$$<{\mathop A\limits_\times}{}_j\cdot
{\mathop A\limits_\bullet}{}^j_\wedge>=
<{\mathop A\limits_\times}{}_j\cdot
\partial_k{\mathop{\cal H}\limits_\times}{}^{jk}_\wedge>=
-2<\partial_{[k}{\mathop A\limits_\times}{}_{j]}\cdot
{\mathop {\cal H}\limits_\times}{}^{jk}_\wedge>
=<{\mathop B\limits_\bullet}{}_{jk}\cdot
\star{\mathop {\cal B}\limits_\bullet}{}_{jk}>=
-<{\mathop B\limits_\bullet}{}_{jk}\cdot
\star\ddag{\mathop B\limits_\bullet}{}_{jk}>.$$

It is sad that the authors of [4] call {\it rest mass} the mass.
Actually, mass is the equivalent of inertia of a body and varies with
speed of the body [5, 6]. \medskip

{\bf 6. The standard electrodynamics spin tensor is zero} \medskip

The energy-momentum tensor $T^{\alpha\gamma}$ and the spin tensor
$\Upsilon^{\alpha\gamma\beta}$ (upsilon) are defined by the following
equalities:
$$ dP^{\alpha} = T^{\alpha\gamma}dV_{\gamma}, \qquad
dS^{\alpha\gamma} = \Upsilon ^{\alpha\gamma\beta}dV_{\beta}
\qquad \alpha,\gamma,\dots = 0,1,2,3.$$
Here infinitesimal 4-momentum $dP^{\alpha}$ and 4-spin
$dS^{\alpha\gamma}$ are observable quantities and $dV_{\gamma}$ is an
3-element. So true definitions of the energy-momentum and spin tensors
do not admit any arbitrariness.

The electrodynamic energy-momentum tensor is the Minkowski tensor.
$$ T^{\alpha\gamma} = - F^{\alpha\nu}F^{\gamma}{}_\nu +
g^{\alpha\gamma}F_{\nu\mu}F^{\nu\mu}/4. $$
Only this tensor satisfies experiments. Only this tensor localizes
energy-momentum. The source of the Minkowski tensor is
$$\partial_\gamma T^{\alpha\gamma} = -F^\alpha_\gamma j^\gamma.$$
The Minkowski tensor is the true electrodynamic energy-momentum tensor.
But a true spin tensor in the electrodynamics is unknown,
$$\Upsilon^{\alpha\gamma\beta}= \mbox{?}$$

In the electrodynamics the variational principle results in a pair of
the canonical tensors: the canonical energy-momentum tensor ${\mathop
T\limits_c}^{\alpha\gamma}$ and the canonical spin tensor
${\mathop\Upsilon\limits_c}^{\alpha\gamma\beta}$ (upsilon):
$$ {\mathop T\limits_c}^{\alpha\gamma} =
- \partial ^{\alpha}A_{\mu}\cdot F^{\gamma\mu} +
g^{\alpha\gamma}F_{\mu\nu}F^{\mu\nu}/4, \qquad
{\mathop\Upsilon\limits_c}^{\alpha\gamma\beta}= -
2A^{[\alpha}F^{\gamma]\beta}.$$

These tensors contradict experience. It is obvious in view of a
asymmetry of the energy-momentum tensor, and it was checked on directly
[2].

An attempt is known to turn the canonical energy-momentum tensor to the
Minkowski tensor by subtraction the Rosenfeld's pair of tensors,
$$ ({\mathop T\limits_R}^{\alpha\gamma}, \,
{\mathop\Upsilon\limits_R}^{\alpha\gamma\beta}) =
(\partial_{\beta}{\mathop\Upsilon\limits_c}^{\{\alpha\gamma\beta\}}/2,
\,{\mathop\Upsilon\limits_c}^{\alpha\gamma\beta}),
\quad \Upsilon ^{\{\alpha\gamma\beta\}} =
\Upsilon ^{\alpha\gamma\beta} - \Upsilon ^{\gamma\beta\alpha} +
\Upsilon ^{\beta\alpha\gamma}, $$
from the canonical pair of tensors.

The Rosenfeld's pair is closed in the sense:
$$\partial_\gamma{\mathop T\limits_R}^{\alpha\gamma}=0,\qquad
\partial_\beta{\mathop\Upsilon\limits_R}^{\alpha\gamma\beta} =
2{\mathop T\limits_R}^{[\alpha\gamma]}.$$
So, the Rosenfeld's pair is closed relative both momentum and spin.
This implies that external sources of the Rosenfeld's pair are zero.

Subtracting the Rosenfeld's pair yields
$$ {\mathop T\limits_c}^{\alpha\gamma} -
{\mathop T\limits_R}^{\alpha\gamma} =
T^{\alpha\gamma} - A^{\alpha}j^{\gamma}, \qquad
{\mathop\Upsilon\limits_c}^{\alpha\gamma\beta} -
{\mathop\Upsilon\limits_R}^{\alpha\gamma\beta} =0. $$

So, the subtraction eliminates the spin tensor and, in the case of
$j^\gamma=0$, yields the Minkowski energy-momentum tensor.

The elimination of the electrodynamic spin tensor provokes a strange
opinion that a circularly polarized plane wave with infinite extent can
have no angular momentum [7, 8], that only a quasiplane wave of finite
transverse extent carries an angular momentum whose direction is along
the direction of propagation. This angular momentum is provided by an
outer region of the wave within which the amplitudes of the electric
$E$ and magnetic $B$ fields are decreasing. These fields have
components parallel to wave vector there, and the energy flow has
components perpendicular to the wave vector. ``This angular momentum is
the spin of the wave'' [9]. Within an inner region the $E$ and $B$
fields are perpendicular to the wave vector, and the energy-momentum
flow is parallel to the wave vector [10].  There is no angular momentum
in the inner region [9].

But let us suppose now that a circularly polarized beam is absorbed by
a round flat target which is divided concentrically into outer and
inner parts. According to the previous reasoning, the inner part of the
target will not perceive a torque. Nevertheless R. Feynman [11] clearly
showed how a circularly polarized plane wave transfers a torque to an
absorbing medium. What is true? And if R. Feynman is right, how one can
express the torque in terms of pondermotive forces?

From our point of view, classical electrodynamics is not complete. The
task is to discover the nonzero spin tensor of electromagnetic field.
\medskip

{\bf 7. Spin tensor of electromagnetic waves} \medskip

For getting the Minkowski tensor from the canonical tensor in the
general case of $j^\gamma\ne0$ we have to subtract
$$\tilde T^{\alpha\gamma}=
{\mathop T\limits_R}^{\alpha\gamma} - A^\alpha j^\gamma$$
from ${\mathop T\limits_c}^{\alpha\gamma}:$
$$ T^{\alpha\gamma} =
{\mathop T\limits_c}^{\alpha\gamma}-
\tilde T^{\alpha\gamma}.$$
What tensor $\tilde\Upsilon^{\alpha\gamma\beta}$ must then we subtract
from ${\mathop\Upsilon\limits_c}^{\alpha\gamma\beta}$ for getting the
true spin tensor?

We suggested that
$$\tilde\Upsilon^{\alpha\gamma\beta} =
{\mathop\Upsilon\limits_c}^{\alpha\gamma\beta} -
2A^{[\alpha}\partial^{\mid\beta\mid}A^{\gamma]}$$
because such a $\tilde\Upsilon^{\alpha\gamma\beta}$
is closed relative to spin:
$$\partial_\beta\tilde\Upsilon^{\alpha\gamma\beta}=
2\tilde T^{[\alpha\gamma]}.$$
This way we obtain an {\it electric} spin tensor:
$${\mathop\Upsilon\limits_e}^{\alpha\gamma\beta}=
{\mathop\Upsilon\limits_c}^{\alpha\gamma\beta}-
\tilde\Upsilon^{\alpha\gamma\beta}=
2A^{[\alpha}\partial^{\mid\beta\mid}A^{\gamma]}.$$

The electromagnetic covariant tensor field $F_{\mu\nu}$ is closed:
$$\partial_{[\alpha} F_{\mu\nu]}=0$$.
But, for waves the conjugate tensor is closed too (we omit wedge in
Sec. 6, 7, 8):
$$\partial_\nu (\star F_{\mu\nu})=\partial_\nu F^{\mu\nu}= j^{\mu}=0.$$
Therefore $F^{\mu\nu}$ has a filling, $\Pi^{\mu\nu\sigma},$
$$\partial_\sigma \Pi^{\mu\nu\sigma} =F^{\mu\nu}.$$
We call $\Pi^{\mu\nu\sigma}$ an {\it electric 3-vector potential}.

The ``Lorentz condition'', $\partial\star(\Pi^{\mu\nu\sigma})=0,$
singles an electric vector potential out from the collection of the
gauge equivalent potentials. We call an electric vector potential the
true potential ${\mathop \Pi\limits_\bullet}{}_{\mu\nu\sigma}$ if
$\partial_{[\lambda}{\mathop\Pi\limits_\bullet}{}_{\mu\nu\sigma]}=0.$

${\mathop \Pi\limits_\bullet} ^\alpha=
\epsilon^{\alpha\mu\nu\sigma}
{\mathop \Pi\limits_\bullet}{}_{\mu\nu\sigma}$ and
${\mathop A\limits_\bullet} ^\alpha$
are equal in rights. So the spin tensor must be symmetric
relative to the magnetic and the electric potentials. Therefore we
suggested that the spin tensor of electromagnetic waves is the sum:
$$\Upsilon^{\alpha\gamma\beta}=
{\mathop\Upsilon\limits_e}^{\alpha\gamma\beta}+
{\mathop\Upsilon\limits_m}^{\alpha\gamma\beta}=
{\mathop A\limits_\bullet} ^{[\alpha}\partial^{\mid\beta\mid}
{\mathop A\limits_\bullet} ^{\gamma]}+
{\mathop \Pi\limits_\bullet} ^{[\alpha}\partial^{\mid\beta\mid}
{\mathop \Pi\limits_\bullet} ^{\gamma]}, \qquad
\partial_\alpha{\mathop A\limits_\bullet} ^\alpha=
\partial_\alpha{\mathop \Pi\limits_\bullet} ^\alpha= 0.\eqno(3)$$
\medskip

{\bf 8. Circularly polarized standing wave} \medskip

Let us consider a circularly polarized plane wave which falls upon a
superconducting $x,y$-plain, and reflects from it, and so a
standing wave forms. The flux density of energy (or volumetric momentum
density) is equal to zero in the wave,
$T^{tz} = \mbox{\bf E}\times \mbox{\bf B} = 0$. But the volumetric
densities of electrical and magnetic energy vary with $z$
in anti-phase. So the total energy density is constant. The momentum
flux density, that is the pressure, is constant too:
$$E^2/2=1-\cos2kz, \quad B^2/2=1+\cos2kz, \quad
T^{tt}=T^{zz}=(E^2+B^2)/2=2.$$

It is interesting to calculate an output of the expression (3) in the
situation. The spin flux density must be zero, $\Upsilon^{xyz}= 0$, and
it is expected that the volumetric spin density consists of electrical
and magnetic parts which are shifted relative to one another. This
result is obtained below.

A circularly polarized plane wave which propagates along $z$-direction
involves the vectors {\bf B}, {\bf E}, {\bf A}, $\Pi$ which lay in
$xy$-plane, and we shall represent them by complex numbers instead of
real parts of complex vectors.
$$\mbox{\bf B}=\{B^x,\ B^y\}\to\ B=B^x+iB^y.$$
Then the product of a complex conjugate number $\overline{E}$ and other
number $B$ is expressed in terms of scalar and vector products of the
corresponding vectors. For example:
$$\overline{E}\cdot B=
(\mbox{\bf E}\cdot\mbox{\bf B})+i(\mbox{\bf E}\times\mbox{\bf B})^z.$$

Since all this vectors do not vary with $x$ and $y$, then
$${\rm curl}\mbox{\bf B}=
\{-\partial_zB^y,\ \partial_zB^x\}\to i\partial_zB,\quad
{\rm curl}^{-1}\to -i\int dz.$$

The angular velocity of all the vectors is $\omega$ and the wave number
along $z$-axis is $k =\omega.$ Therefore
$$\mbox{\bf B}\to B_{01}e^{i\omega(t-z)}\
\mbox{or, for a reflected wave, } B_{02}e^{i\omega(t+z)},$$
$$\partial_t\to i\omega,\quad\partial_z\to\mp i\omega,\quad
{\rm curl}\to\pm\omega,\quad{\rm curl}^{-1}\to\pm1/\omega.$$

If $z = 0$ at the superconducting $x,y$-plain, then the falling and
reflected waves are recorded as
$$B_1=e^{i\omega(t-z)},\quad E_1=-ie^{i\omega(t-z)},\quad
B_2=e^{i\omega(t+z)},\quad E_2=ie^{i\omega(t+z)}.$$
The complex amplitudes are equal here:
$B_{01}=B_{02}=1,\ E_{01}=-i,\ E_{02}=i.$

Since
$\mbox{\bf A}={\rm curl}^{-1}\mbox{\bf B},\
\Pi={\rm curl}^{-1}\mbox{\bf E},$
the other complex amplitudes are received by a simple calculation (time
derivative is designated by a point):
$$A_{01}=1/\omega,\ \dot A_{01}=i,\
\Pi_{01}=-i/\omega,\ \dot\Pi_{01}=1,\
A_{02}=-1/\omega,\ \dot A_{02}=-i,\
\Pi_{02}=-i/\omega,\ \dot\Pi_{02}=1.$$

Now we calculate the electric and magnetic parts of the volumetric spin
density.
\begin{eqnarray*}
{\mathop\Upsilon\limits_e}^{xyt}&=&
(\mbox{\bf A}\times\dot{\mbox{\bf A}})/2=
\Im(\overline{(A_1+A_2)}\cdot({\dot A}_1+{\dot A}_2))/2\\
&=&\Im((e^{-i\omega(t-z)}-e^{-i\omega(t+z)})i
(e^{i\omega(t-z)}-e^{i\omega(t+z)}))/2\omega=
(1-\cos 2\omega z)/\omega,
\end{eqnarray*}
$${\mathop\Upsilon\limits_m}^{xyt}=
(\Pi\times\dot\Pi)/2=
\Im(\overline{(\Pi_1+\Pi_2)}\cdot({\dot\Pi}_1+{\dot\Pi}_2))/2=
(1+\cos 2\omega z)/\omega,$$
$$\Upsilon^{xyt}=
{\mathop\Upsilon\limits_e}^{xyt}+
{\mathop\Upsilon\limits_m}^{xyt}=2/\omega.$$

So, the terms which oscillate along $z$-axis cancel out.
It is easy to calculate that spin flux is equal to zero (the prime
denote the derivative with respect to z):
$${\mathop\Upsilon\limits_e}^{xyz}=
-(\mbox{\bf A}\times{\mbox{\bf A}}')/2=0, \quad
{\mathop\Upsilon\limits_m}^{xyz}=
-(\Pi\times\Pi')/2=0.$$
\medskip

{\bf 9. A new anthropic principle} \medskip

This section contains an important idea that is not
concerned with the above-stated topic.  I am forced to
insert this text in the paper because T. Schwander and Mr.
Kristrun removed it from the arXiv twice.

The fact is that the mankind is lonely in the
universe. We listen to the universe, but we
hear nobody. We shout in the universe, but nobody
answers. People have thought up UFOs and panspermia
because of melancholy of the loneliness.
The believing scientist Blez Paskal' wrote:
``Eternal silence of these boundless spaces horrifies
me. Silence is the greatest of all
persecutions. Saints never were silent.'' However
scientists should not be grieved. They
must reply why we are lonely.

     An answer to the question, why we are lonely,
naturally, is connected to a question
why we exist. As is known, we exist due to physical
laws and values of physical
constants are favorable for life. However, according
to the anthropic principle, our
universe is fine-tuned. A little change of physical
constants may make the universe is
unsuitable for life. In other words, the area of
values of the constants permitting life is
extremely small. We shall name this area the
anthropic area. Why constants have got in
the anthropic area?

     Reasons of the realization of such improbable
values of the constants in our universe
were discussed at representative conference
``Anthropic arguments in fundamental
physics and cosmology'' which was held in Cambridge
from 30 August -- 1 September
(See ``Physics World'' October 2001, p. 23). The
basic idea was multiple universes. This
means that  an enormous or infinite amount of
universes with various physical laws and
various physical constants can exist. But one can
observe only  universe which permits
existence of an observer, i. e. an anthropic universe.
So, the small probability of an
anthropic universe is of no importance for the
observer. Nevertheless the reason of our
loneliness was not discussed at the conference.

      Meanwhile it is obvious that among anthropic
universes there are universes which are
more or less favorable for life. And if the
probability of an anthropic universes is very
small against the background of all universes, it is
natural to expect that the probability
of an especially favorable for life universe will be
very small against the background of
universes that are simply favorable for life and,
especially, that are rather adverse,
marginal universes adjoining to non-anthropic
universes.

      In such adverse universes which take place at an
edge of anthropicness, life should
arise extremely seldom, only in a case of confluence
of many favorable circumstances.
But such universes are the most probable among
anthropic universes. Therefore it is not
necessary to be surprised that the universe in which
we have pleasure to stay is stingy
with the organization of life.

      It is possible to exemplify this argument.
Outstanding astrophysicist I. S. Shklovsky
explained our loneliness. He wrote: ``Practically all
stars such as our Sun are parts of
double (or multiple) systems. Life may not develop
in such systems as the temperature
of surfaces of their hypothetical planets should vary
in inadmissible wide limits. It looks
as if our Sun, a strange single star surrounded with
family of planets,  is an exception in
the world of stars''.

     Our idea is the following. Let us admit that it
is possible such a change of the
constants which will increase the percentage of
single stars in the universe and will make
life more widespread. But such universe would be less
probable than ours. Its physical
constants should get in a very small privileged part
of the antropic area.

     So, probably, there are no bases to admire with
successful values of the present
physical constants. Since our loneliness in the
universe the constants are not so
successful. And this was necessary to expect. It was
necessary to expect that we had
appeared in one of the most probable anthropic
universes, in such which is the worst
adapted to life by virtue of its greatest
probability. We exist, and it is pleasant, but we
exist, probably, in loneliness, on the edge of life
in a metauniverse sense. A new anthropic
principle can be formulated as follows. It is the
most probable to observe such universe
in which life is an extremely rare phenomenon. We
have just that very case. \medskip

\centerline{\bf Note } \medskip

This paper matter is contained in the following papers which were
submitted to the following journals (all the journals rejected or
ignored all the papers):

``Electromagnetism in terms of sources and generations of fields''
{\it Physics - Uspekhi} (13 June, 1995).

``Electromagnetism: sources, generations, boundaries'', ``Spin tensor
of electromagnetic fields'' {\it J. Experimental \& Theor. Phys. Lett.}
(14 May, 1998).

``Spin tensor of electromagnetic fields'' {\it J. Experimental \&
Theor. Phys.} (27 Jan. 1999), {\it Theor. Math. Phys.} (29 Apr. 1999),
{\it Rus. Phys. J.} (18 May, 1999).

``Are spin and orbital momentum the same quantity?'' {\it J.
Experimental \& Theor. Phys.}, {\it Physics - Uspekhi} (25 Feb. 1999),
{\it Rus. Phys. J.} (15 Oct. 1999).

``Energy-momentum and spin tensors problems in the electromagnetism''
{\it J. Experimental \& Theor. Phys.} (13 Apr. 2000), {\it Physics -
Uspekhi} (12 Jan. 2000), {\it Rus. Phys. J.} (1 March. 2000), {\it
Theor. Math. Phys.} (17 Feb. 2000).

``Angular momentum distribution of the rotating dipole field'' {\it J.
Experimental \& Theor. Phys.}, {\it Rus. Phys. J.} (25 May, 2000), {\it
Theor. Math. Phys.} (29 May, 2000). {\it Physics - Uspekhi} (31 May,
2000).

``Electromagnetism in terms of boundaries and generations of
differential forms'' {\it Physics - Uspekhi} (4 Oct. 2000).

``Tubes of force and bisurfaces in the electromagnetism'' {\it Physics
- Uspekhi} (28 March, 2001), {\it Rus. Phys. J.} (26 Apr. 2001).

``Violation of the gauge equivalence'' {\it Theor. Math. Phys.}, {\it J.
Experimental \& Theor. Phys.} (16 May, 2001), {\it Rus. Phys. J.} (31
May, 2001), {\it Foundations of Physics} (28 May, 2001).

The subject matter of this paper had been partially published [1--3,
12, 13]. \medskip

\centerline{\bf References}

1. Khrapko R. I., ``Spin density of electromagnetic waves'',\\
http://www.mai.ru/projects/mai\_works/index.htm

2. Khrapko R. I., ``True energy-momentum tensors are unique.
Electrodynamics spin tensor is not zero'', physics/0102084.

3. Khrapko R. I., ``Tubes of force and bisurfaces in
electromagnetism'', \\http://www.mai.ru/projects/mai\_works/index.htm

4. Gubarev F. V., Stodolsky L., Zakharov V. I., ``On the Significance
of the Vector Potential Squared'', Phys. Rev. Lett., {\bf 86}, 2220
(2001)

5. Khrapko R. I., ``What is mass?'', physics/0103051, Physics - Uspekhi
{\bf 43} (12) 1267 (2000)

6. Khrapko R. I., ``Rest mass or inertial mass?'', physics/0103008.

7. Heitler W., The Quantum Theory of Radiation (Clarendon, Oxford,
1954), p. 401.

8. Simmonds J. W., Guttmann M. J., States, Waves and Photons (Addison-
Wesley, Reading, MA, 1970).

9. Ohanian H. C., ``What is spin?'' Amer. J. Phys. {\bf 54}, 500,
(1986).

10. Jackson J. D., Classical Electrodynamics (John Wiley, New York,
1962), p. 201.

11. Feynman R. P., {\it et al}. The Feynman Lectures On Physics
(Addison-Wesley, London, 1965), v. 3, p. 17-10.

12. Khrapko R. I., ``True energy-momentum and spin tensors are
unique.'' {\it In} Abstracts of 10-th Russian Gravitational Conference
(Moscow, 1999), p. 47.

13. Khrapko R. I., ``Does plane wave not carry a spin?'' Amer. J. Phys.
{\bf 69}, 405, (2001).
\end{document}